\documentclass[aps,prl,reprint,twocolumn,superscriptaddress,showpacs]{revtex4-2}

\usepackage{graphicx}
\usepackage{mathrsfs}
\usepackage{bm}
\usepackage{amsmath}
\usepackage{dcolumn}
\usepackage{dsfont}
\usepackage{amssymb}
\usepackage{tabularx}
\usepackage{array}
\usepackage{float}
\usepackage{color}
\usepackage{epstopdf}
\usepackage{mathrsfs}
\usepackage[T1]{fontenc}
\usepackage[colorlinks, linkcolor=blue,anchorcolor=blue,citecolor=blue,urlcolor=blue]{hyperref}

\usepackage[mathlines]{lineno}
\usepackage{bbold}
\usepackage[T1]{fontenc}
\usepackage{multirow}

\usepackage{gensymb}
\graphicspath{{./}{figs/}}

\begin{document}

\preprint{\textit{Preprint: \today, \now.}}

\title{Electronic correlations in the normal state of kagome superconductor KV$_3$Sb$_5$}

\author{Jianzhou Zhao}
\email{jzzhao@swust.edu.cn}
\affiliation{Co-Innovation Center for New Energetic Materials, Southwest University of Science and Technology, Mianyang 621010, China}
\affiliation{Research Laboratory for Quantum Materials, Singapore University of Technology and Design, Singapore 487372, Singapore}

\author{Weikang Wu}
\affiliation{Research Laboratory for Quantum Materials, Singapore University of Technology and Design, Singapore 487372, Singapore}
\affiliation{Division of Physics and Applied Physics, School of Physical and Mathematical Sciences, Nanyang Technological University, Singapore 637371, Singapore}

\author{Yilin Wang}
\email{yilinwang@ustc.edu.cn}
\affiliation{Hefei National Laboratory for Physical Sciences at Microscale, University of Science and Technology of China, Hefei, Anhui 230026, China}

\author{Shengyuan A. Yang}
\affiliation{Research Laboratory for Quantum Materials, Singapore University of Technology and Design, Singapore 487372, Singapore}

%
\begin{abstract}
	Recently, intensive studies have revealed fascinating physics, such as charge density wave  and superconducting states, in the newly synthesized kagome-lattice materials $A$V$_3$Sb$_5$ ($A$=K, Rb, Cs). Despite the rapid progress, fundamental aspects like the magnetic properties and electronic correlations in these materials have not been clearly understood yet.
Here, based on the density functional theory plus the single-site dynamical mean-field theory calculations, we investigate the correlated electronic structure and the magnetic properties of the KV$_3$Sb$_5$ family materials in the normal state.
We show that these materials are good metals with weak local correlations. The obtained Pauli-like paramagnetism and the absence of local moments are consistent with recent experiment. We reveal that the band crossings around the Fermi level form three groups of nodal lines protected by the spacetime inversion symmetry, each carrying a quantized $\pi$ Berry phase. Our result suggests that the local correlation strength in these materials appears to be too weak to generate unconventional superconductivity, and non-local electronic correlation might be crucial in this kagome system.
\end{abstract}


\maketitle

The kagome lattice has attracted great interest in condensed matter physics research. As a prototype lattice with strong geometric frustration, the kagome lattice has been extensively studied in quantum magnetism~\cite{Pollmann.2008, Shores.2005} and was proposed as host for a quantum spin liquid state~\cite{Yan.2011, Han.2012,Han.2012ss,Fu.2015}. Itinerant electrons on a kagome lattice can realize a special band structure with Dirac cones and flat band~\cite{Ghimire.2020,Ye.2018gb8}. Further incorporating electron interaction effects, a rich variety of exotic effects have been predicted on the kagome lattice, such as charge bond order~\cite{Brien.2010,Pollmann.2014}, spin or charge density waves (SDW or CDW)~\cite{Wang.2012fkd}, charge fractionalization~\cite{Ruegg.20115ja}, topological insulating state~\cite{Guo.2009,Wen.2010}, and superconductivity~\cite{Kiesel.2012,Kiesel.2013}. Driven by these predictions, real materials that contain a kagome lattice have been actively explored~\cite{Ye.2018gb8,Wang.2016nm,Liu.2018c9p,Yin.2018,Yin.2019pk,Wang.2018,Xu.2020,Yin.2020}.

Recently, a novel family of kagome materials $A$V$_3$Sb$_5$ ($A$ = K, Rb, Cs) were synthesized~\cite{Ortiz.2019}. These materials share the same layered structure, which contains active layers of a V kagome lattice. All the three compounds exhibit superconductivity at low temperature, with $T_c=0.93$, 0.92 and 2.5 K, for KV$_3$Sb$_5$~\cite{Ortiz.2021}, RbV$_3$Sb$_5$~\cite{Yin.2021} and CsV$_3$Sb$_5$~\cite{Ortiz.202087}, respectively.
Besides superconductivity, a CDW instability was observed at $T^* \sim 80$ to 100 K~\cite{Ortiz.2019,Ortiz.202087,Jiang.2020,Ortiz.2021,Yin.2021,Li.2021kl,Liang.2021,Tan.2021}. Interestingly,
in the normal state above the CDW transition, these materials were revealed to be $\mathbb{Z}_2$ topological metals~\cite{Ortiz.202087}. The angle-resolved photoemission spectroscopy (ARPES) experiment reported multiple Dirac points near the Fermi level. This is also supported by the Shubnikov-de Haas oscillation result that indicates highly dispersion low-energy bands with rather low effective mass~\cite{Yin.2021}. It was speculated that the Dirac electrons could play an important role in the observed large non-spontaneous anomalous Hall response~\cite{Yang.2020zj,Yu.2021}.

Despite the rapid progress, there are still many puzzles on these materials to be addressed. For instance, the recent susceptibility measurement and $\mu$SR measurement both indicate the absence of local magnetic moments~\cite{Ortiz.2021,Kenney.2020}, which differs from the expectation from simple valence counting and appears to be in conflict with the observed extremely large anomalous Hall response. Meanwhile, the nature of the superconductivity in these materials is still under debate~\cite{Duan.2021,Zhang.2021,Tan.2021,Li.2021kl,Yu.2021,Zhao.2021,Chen.2021,Chen.2021tlz,Chen.2021i7l,Yin.2021,Du.2021}. Some of the recent theoretical and experimental works suggested that the superconductivity and CDW might be unconventional~\cite{Li.2021kl,Yu.2021,Zhao.2021,Chen.2021,Chen.2021tlz}, which hints at important electron correlation effects. However, an understanding of the correlation strength in these materials is still lacking.

In this work, we investigate the correlated electronic structure and magnetic properties of KV$_3$Sb$_5$, using the combination of density functional theory (DFT) and dynamical mean field theory (DMFT) calculations~\cite{Georges.1996, Kotliar.2006}.
By considering the on-site Coulomb interaction, we show that the KV$_3$Sb$_5$ family materials are weakly correlated metals in the normal state.
Our obtained temperature independence of Pauli-like magnetic susceptibility behavior is consistent with experimental result~\cite{Ortiz.2021,Kenney.2020}, confirming the absence of local moments. From the atomic configuration analysis and the calculated hybridization function, we attribute this result to the strong $p$-$d$ hybridization that leads to the delocalization of V-$d$ electrons. The low-energy bands are only weakly affected by the on-site interaction. We analyze the linear band crossings near the Fermi level, and show that they are actually not isolated Dirac points, instead, they belong to three groups of nodal lines in the Brillouin zone (BZ), protected by the spacetime inversion symmetry in the absence of spin-orbit coupling (SOC). One group (three rings) is centered around the $M$ point, the other two vertically traverse the BZ. Our result indicates that the local electron-electron correlation alone is not sufficient to account for unconventional superconductivity (if it indeed exists); nonlocal correlations might be required for such an effect. This also echoes with recent works on the CDW state of these materials, which suggest that nonlocal correlations play an important role in its formation~\cite{Li.2021kl}.



\begin{table}[ht]
    \caption{\label{tab:lattice} Lattice parameters and atom positions for KV$_3$Sb$_5$ used in our DFT+DMFT calculation.}
    \begin{ruledtabular}
        \begin{tabular}{cccccc}
            $a$(\AA)     & $b$(\AA)     & $c$(\AA)       & $\alpha$ & $\beta$ & $\gamma$ \\
            5.48213      & 5.48213      & 8.95802        & 90.0     & 90.0    & 120.0    \\ \hline\hline
            \multirow[b]{2}{*}{1}       & Site         & Wyckoff symbol & $x$      & $y$     & $z$      \\ \cline{2-6}
                         & K            & 1$a$           & 0        & 0       & 0        \\
            2            & V            & 3$g$           & 1/2      & 1/2     & 1/2      \\
            3            & Sb$_1$       & 4$h$           & 2/3      & 1/3     & 0.753    \\
            4            & Sb$_2$       & 1$b$           & 0        & 0       & 1/2      \\
        \end{tabular}
    \end{ruledtabular}
\end{table}

\begin{figure}[h]
	\centering
	\includegraphics[width = 0.5\textwidth]{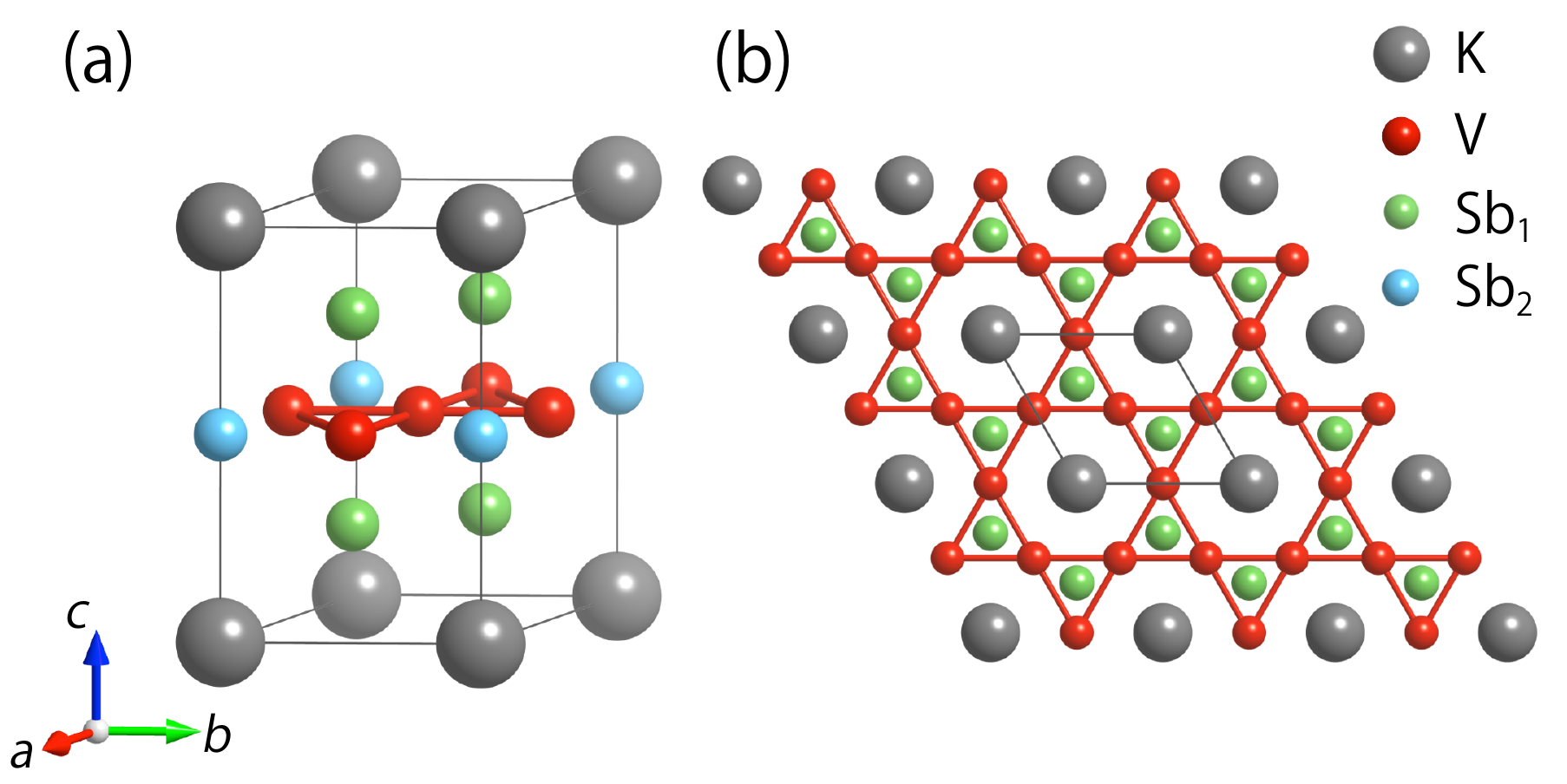}
	\caption{(a) Crystal structure of KV$_3$Sb$_5$.  Here shows a unit cell. (b) Top view of KV$_3$Sb$_5$. The red lines show the kagome lattice of V. Sb atoms in the V plane are labeled as Sb$_2$. Other Sb atoms are labeled as Sb$_1$.}
	\label{fig:lattce}
\end{figure}

{\color{blue}\textit{Method.}} We perform fully charge self-consistent DFT+DMFT calculations using the {\sc edmftf} package~\cite{Haule.2010}, based on the full-potential linear augmented plane-wave method implemented in the {\sc wien2k} code~\cite{wien2k,Blaha.2020}.
The $k$-point mesh for the Brillouin zone integration is $17\times 17\times 10$, and the plane wave cut-off $K_{max}$ is given by $R_{MT}\times K_{max}=8.0$. We employed the generalized gradient approximation (GGA) with the Perdew-Burke-Ernzerhof (PBE) realization~\cite{Perdew.1996} as exchange-correlation functional. The atomic sphere $R_{MT}$ are 2.50, 2.50, 2.65 a.u. for K, V and Sb, respectively.
We use projectors with energy window from $-10$ to 10 eV relative to the Fermi level to construct V-$3d$ local orbitals.
A rotationally invariant form of local on-site Coulomb interaction Hamiltonian parameterized by Hubbard $U$ and Hund's coupling $J_H$ is applied on all five V-$3d$ orbitals. We choose $U=5.0$ eV and $J_H=0.7$ eV in this work, which are typical values, as used previously for SrVO$_3$~\cite{Pavarini.2003,Taranto.2013}, V$_2$O$_3$~\cite{Held.20015,Held.2006} and VO$_2$~\cite{Laad.2006}. The impurity problem is solved by the hybridization expansion version of the continuous time quantum Monte Carlo (CTQMC) solver~\cite{Gull.2011}.
We choose an ``exact'' double counting scheme developed by Haule~\cite{Haule.201509p}, in which the Coulomb repulsion in real space is screened by the combination of Yukawa and dielectric functions. The self-energy on real frequency is obtained by the analytical continuation method of maximum entropy~\cite{Haule.2010}.
The effective mass enhancement by correlations is defined by $m^*/m_{\text{DFT}} = 1/\mathcal{Z}$, where $\mathcal{Z}$ is the quasi-particle weight. To avoid large error-bar in analytic continuation, we directly obtain $\mathcal{Z}=1-\frac{\partial\text{Im}\Sigma(\mathrm{i}\omega_n)}{\partial\omega_n}\big|_{\omega_n\rightarrow 0^+}$ from the polynormial fit to self-energies on the first ten Matsubara frequencies. We have tested that SOC has negligible effects on the low-energy band structure, so it is omitted in the calculation.

{\color{blue}\textit{Lattice structure.}} The crystal structure for KV$_3$Sb$_5$ is shown in Fig.~\ref{fig:lattce}. The material has a layered structure, with $P6/mmm$ space group. The most important motif is the kagome lattice formed by the V atoms. The kagome plane is sandwiched by two layers of Sb$_1$ atoms, each consisting of a honeycomb lattice. Within the same lattice plane, the Sb$_2$ atoms fill the large voids of the kagome lattice.  The low-energy states are dominated by these three atomic layers, which form a V-Sb slab. The alkali atoms fill the region between the V-Sb slabs, and they mainly play the role of electron donors. It follows that the three materials in the family should exhibit very similar electronic properties, which is confirmed by experiments and also by our calculation. Hence, we will mainly focus on KV$_3$Sb$_5$ in the following discussion.

We adopt the experimental lattice parameters in the calculation~\cite{Ortiz.2019}. The parameters for KV$_3$Sb$_5$ is shown in Table~\ref{tab:lattice}. The values for the other two members are given in the Supplemental Material (SM)~\cite{suppl}.


\begin{table}[ht]
	\caption{\label{tab:correlation} The orbital-resolved V-$3d$ occupation $n_d$ and effective mass-enhancement $m^*/m_{\text{DFT}}$ for KV$_3$Sb$_5$, RbV$_3$Sb$_5$ and CsV$_3$Sb$_5$ obtained from DFT+DMFT calculation at $T=300$ K.}
	\begin{ruledtabular}
	\begin{tabular}{cc|ccccc}
	    &  &	$d_{z^2}$ & $d_{x^2-y^2}$ & $d_{xz}$ & $d_{yz}$ & $d_{xy}$ \\
		\hline
		\multirow{2}{*}{KV$_3$Sb$_5$}
		& $n_d=3.169$  & 0.678 & 0.526 & 0.605 & 0.575 & 0.785 \\
		& $m^*/m_{\text{DFT}}$  & 1.354 & 1.284 & 1.442 & 1.308 & 1.340 \\
		\hline
		\multirow{2}{*}{RbV$_3$Sb$_5$}
		& $n_d=3.112$  & 0.683 & 0.540 & 0.614 & 0.484 & 0.791 \\
		& $m^*/m_{\text{DFT}}$  & 1.356 & 1.282 & 1.445 & 1.308 & 1.342 \\
		\hline
		\multirow{2}{*}{CsV$_3$Sb$_5$}
		& $n_d=3.206$  & 0.686 & 0.532 & 0.606 & 0.591 & 0.791 \\
		& $m^*/m_{\text{DFT}}$  & 1.351 & 1.292 & 1.446 & 1.313 & 1.347 \\
	\end{tabular}
	\end{ruledtabular}
\end{table}

{\color{blue}\textit{Orbital occupancy and mass enhancement.}} Let us first consider the occupation $n_d$ of V-$3d$ orbitals in  KV$_3$Sb$_5$. From our DFT+DMFT calculations, the $3d$ occupancy of vanadium is about 3.169. The orbital resolved occupancy number ranges from 0.526 to 0.785 with $d_{x^2-y^2}$ ($d_{xy}$) as the least (most) occupied orbitals. These numbers are listed in Table~\ref{tab:correlation}.
The results for the other two members are quite similar, with $n_d=3.112$ for RbV$_3$Sb$_5$ and 3.206 for CsV$_3$Sb$_5$.

Then we consider the mass enhancement $m^*/m_\text{DFT}=1/\mathcal{Z}$, which is widely used to characterize the strength of electronic correlations. This ratio is unity for an uncorrelated normal metal, and is much larger than unity for a strongly correlated system (e.g., $m^*/m_\text{DFT}\approx 7$ in iron-based superconductor FeTe~\cite{Yin.2011pob,Tamai.2010}). Our calculation results for KV$_3$Sb$_5$ are listed in Table~\ref{tab:correlation}. One observes that the mass enhancement of the V-$3d$ eletrons is quite weak, range from 1.284 ($d_{x^2-y^2}$) to 1.442 ($d_{xz}$).
The other two members RbV$_3$Sb$_5$ and CsV$_3$Sb$_5$ have similar values. This indicates that the KV$_3$Sb$_5$ family materials are weakly correlated metals in their normal states.
Our DFT+DMFT results are consistent with the experimental observation of low effective mass and highly dispersive bands in KV$_3$Sb$_5$~\cite{Yang.2020zj} and RbV$_3$Sb$_5$~\cite{Yin.2021}



\begin{figure}[ht]
	\centering
	\includegraphics[width = 0.50\textwidth]{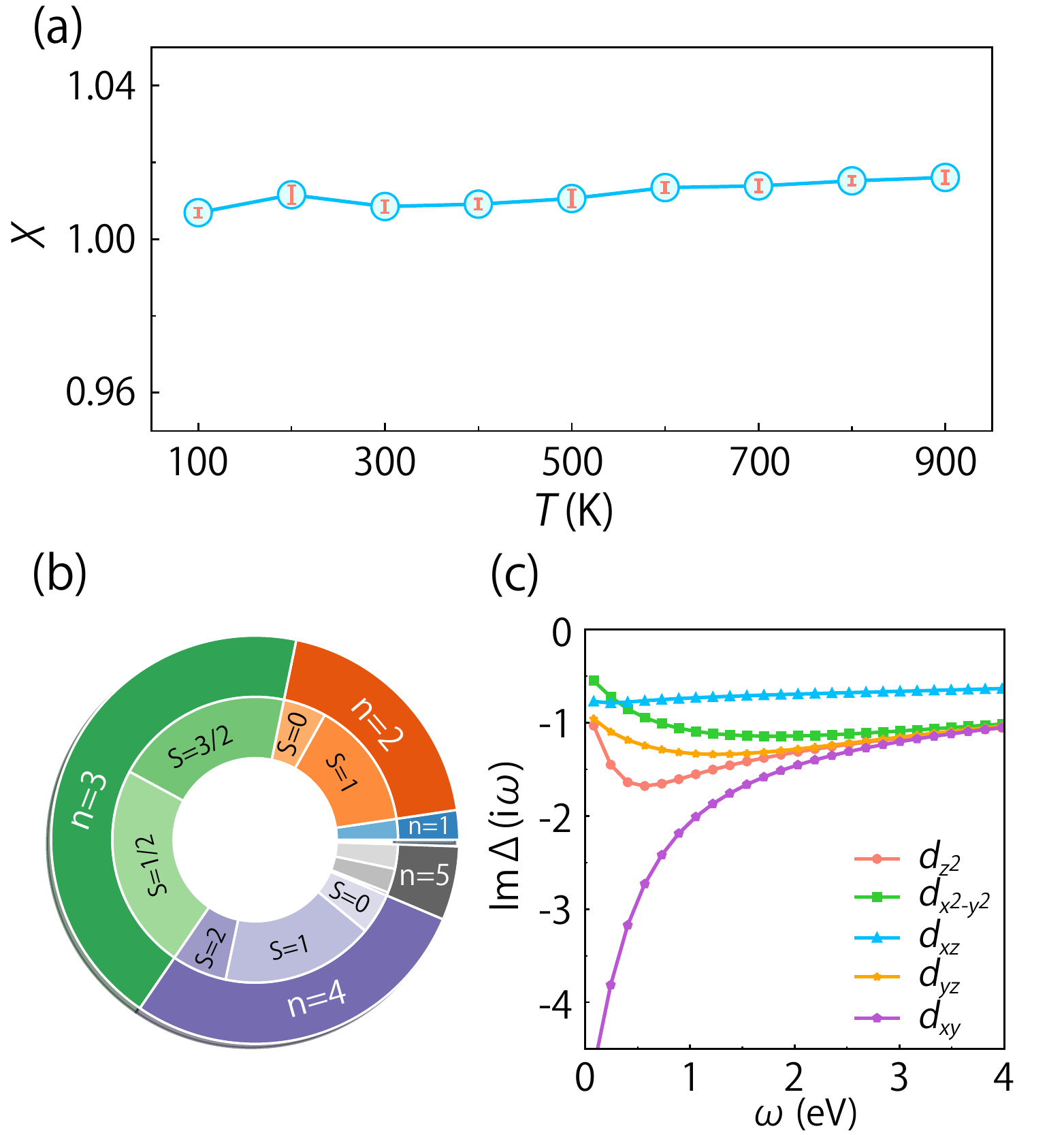}
	\caption{(a) The calculated spin susceptibility $\chi$ as a function of temperature $T$. The error bar indicates the fluctuation of the results for the last five charge self-consistent iterations. (b) The valence histograms of the V-$3d$ shell obtained by CTQMC. (c) Imaginary part of the hybridization function on Matasubara frequencies at $T=300$ K. }
	\label{fig:chi}
\end{figure}

{\color{blue}\textit{Spin susceptibility.}} To investigate the magnetic properties, we calculate the static local spin susceptibility, defined as
\begin{equation}\label{eqn:chi}
\chi=\int^\beta_0{\langle S_z(\tau)S_z(0)\rangle}\mathrm{d}\tau
\end{equation}
by the CTQMC approach. Here, $\beta=1/(k_B T)$, and $\tau$ is the imaginary time. The calculation is done for the normal state in the temperature range from 100 K to 900 K. The obtained $\chi$ versus $T$ result is presented in Fig.~\ref{fig:chi}(a). Clearly, the susceptibility exhibits a paramagnetism with an almost flat line independent of $T$ in the normal state. For good metals, this behavior indicates the dominance of Pauli paramagnetic response from itinerant electrons and the absence of local moments (local moments instead would give a $T$ dependence typically following the Curie-Weiss law). Our result agrees well with the recent magnetic measurements on KV$_3$Sb$_5$~\cite{Ortiz.2021,Kenney.2020}.

To further shed light on this result, in Fig.~\ref{fig:chi}(b), we present the probability distribution for the different atomic configurations for the V-3$d$ shell.
The DMFT atomic basis is constructed from the five $d$ orbitals with the size of $\sum_n{C^n_{10}}=848$ for 7 different occupancies with $n=0,1,\ldots,6$.
One finds that the $n=3$ state has the highest probability with $43.7\%$, followed by $n=4$ with $28.2\%$ and $n=2$ with $19.4\%$.
Within the dominant $n=3$ state, probability for the low spin state is $23.4 \%$, slightly higher than that for the high spin state $S=3/2$ ($\sim 20.3\%$). Note that in the atomic limit, the configure with occupancy $n_d=3$ should typically favor the high spin state according to the Hund's rule. Here,
the close competition between different occupancy and spin states indicate a strong charge and spin fluctuation in the normal state of KV$_3$Sb$_5$.
This can be attributed to  the strong hybridization between V-$3d$ and Sb-$5p$ orbitals. In Fig.~\ref{fig:chi}(c), we plot the hybridization function for the V-$3d$ orbitals, from which one indeed observes that all orbitals are strongly delocalized at low frequency.


\begin{figure}[!ht]
	\centering
	\includegraphics[width = 0.5\textwidth]{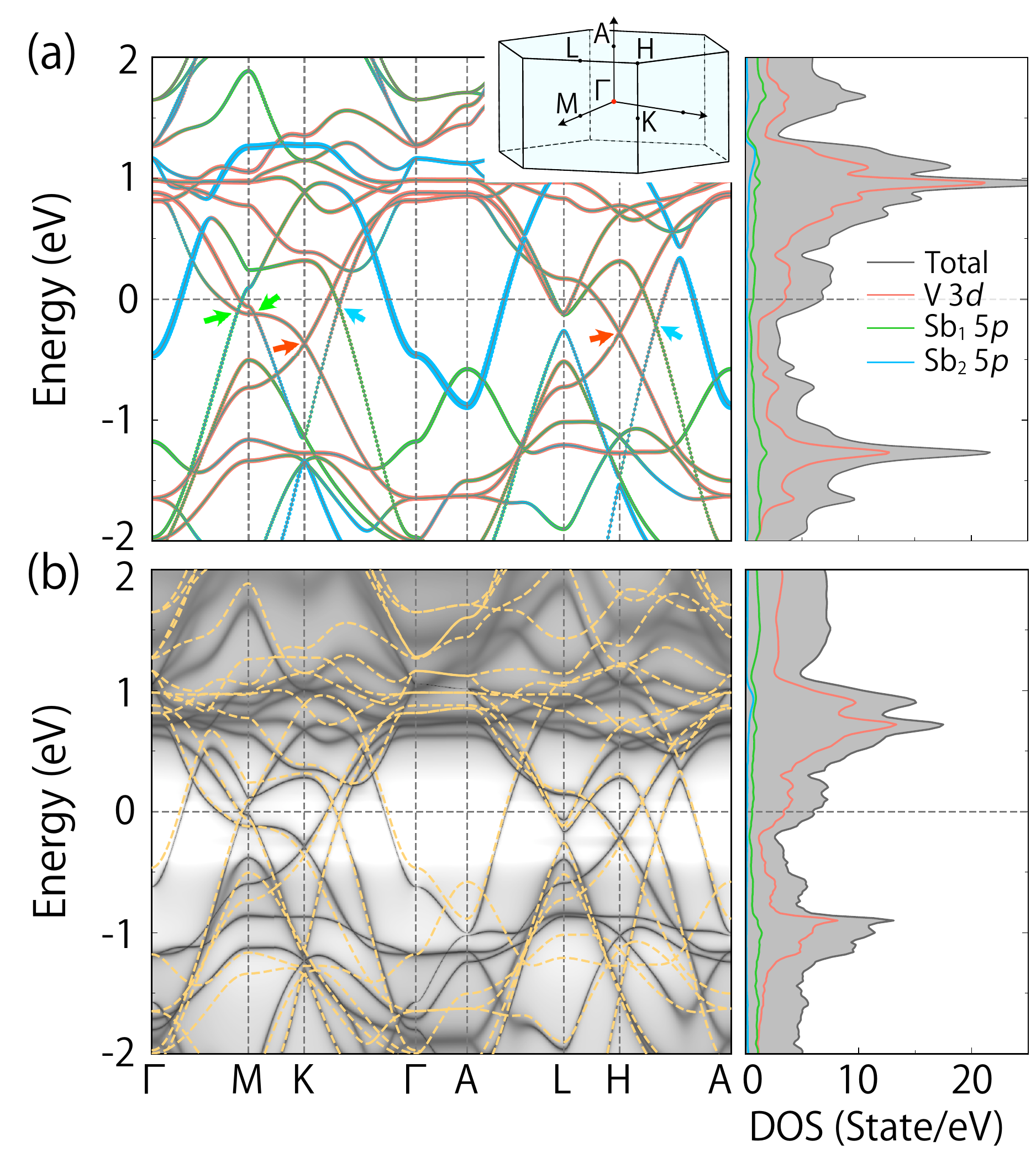}
	\caption{ (a) DFT band structure and DOS of KV$_3$Sb$_5$. The weight of V-$3d$, Sb$_1$ and Sb$_2$ $5p$ orbitals are indicated by red, green and blue colors, respectively. (b) $k$-resolved spectral function obtained by DFT+DMFT at $T=300$ K. The DFT band structure is plotted with gold dashed lines for reference. }
	\label{fig:band}
\end{figure}

{\color{blue}\textit{Correlated electronic structure.}}  The DFT band structure and density of states (DOS) are shown in Fig.~\ref{fig:band} (a). One observes that the bands within 2 eV around the Fermi level are dominated by the V-$3d$ and Sb-$5p$ states.
There is a large peak in DOS around 1 eV with V-$3d$ character, which could be attributed to the flat band featured by the kagome network. There are several linear band crossings around the Fermi level. And a band with V-$3d$ character exhibits a van Hove singularity at the $M$ point around -63 meV. Meanwhile, a highly dispersive band with Sb$_2$-$5p$ character forms two large electron pockets at $\Gamma$ and $A$ points.

As a comparison, the correlated electronic spectral function from our DFT+DMFT calculation is shown in Fig.~\ref{fig:band}(b). One can see that the low-energy bands are not significantly affected by the electronic correlations (the DFT bands are plotted with gold dashed lines in this figure for reference). The bands with V-$3d$ characters between -1 and 1 eV are only slightly pushed above by the renormalization effect introduced by the correlation. The quasiparticle spectrum maintains a good coherence, indicating a negligible imaginary part of self energy and long quasiparticle lifetime. Overall, our result shows that KV$_3$Sb$_5$ is a weakly correlated metal.

\begin{figure}[ht]
	\centering
	\includegraphics[width = 0.5\textwidth]{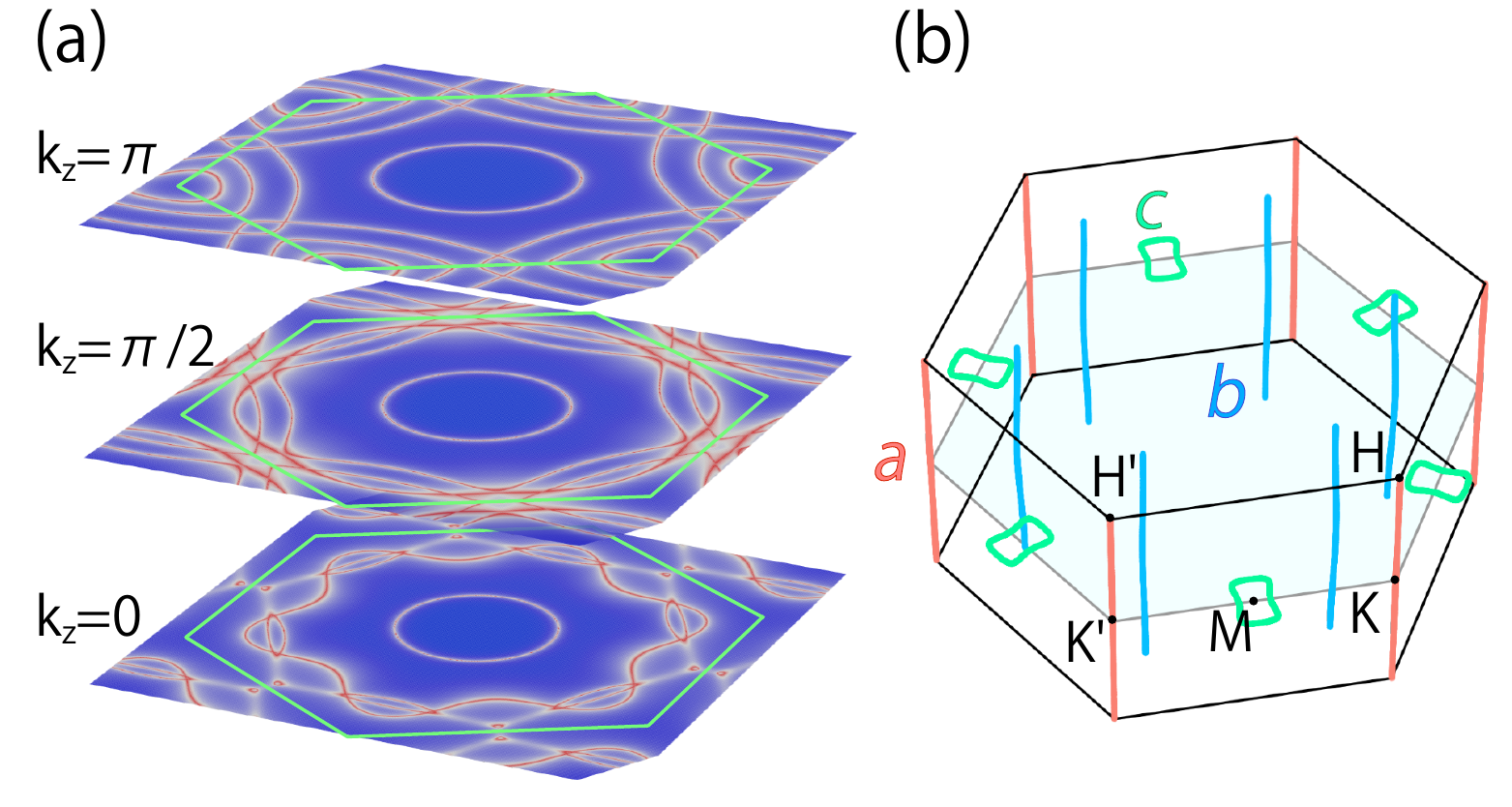}
	\caption{(a) Fermi surface plot for $k_z=0$, $k_z=\pi/2$ and $k_z=\pi$ planes by DFT+DMFT at $T=300$ K. The BZ boundary is shown in green lines. (b) Three types of nodal lines $a$ (red), $b$ (blue), and $c$ (green) in the BZ. Points on these lines are indicated with colored arrows in Fig.~\ref{fig:band}(a). }
	\label{fig:nodal}
\end{figure}

In Fig.~\ref{fig:nodal}(a), we plot the DFT+DMFT Fermi surface contours in the $k_z=0$, $\pi/2$ and $\pi$ slices of the BZ.
The most obvious feature is a highly two-dimensional (2D) electron pocket at the center of BZ, which has been observed in the previous ARPES experiment on CsV$_3$Sb$_5$~\cite{Ortiz.202087}.
According to our calculation~\cite{suppl}, this 2D electron pocket is contributed mainly by the $p_z$ orbitals on Sb$_2$ atoms.
A Fermi surface with hexagonal petal shape is located close to the BZ boundary in the $k_z=0$ plane, which is mainly from the V-$3d_{xy}$ states. And six round pockets with V-$3d_{xz/yz}$ character appear close to the $K$ and $K'$ points.

{\color{blue}\textit{Topological nodal lines.}} The linear band crossings observed in Fig.~\ref{fig:band}(a) around Fermi level were previously interpreted as Dirac points and as a feature of the kagome lattice~\cite{Ortiz.202087,Ortiz.2021}. Nevertheless, we note the following two points. First, the system preserved both inversion $\mathcal{P}$ and time reversal $\mathcal{T}$ symmetries. Under the combined $\mathcal{PT}$ symmetry, an isolated (twofold) Dirac point cannot exist in a 3D system~\cite{Weng.201597g}. Second, the Dirac points for the standard kagome model appear at the high-symmetry $K$ and $K'$ points. However, in Fig.~\ref{fig:band}(a), linear crossings also appear at other locations, such as $\Gamma$-$M$, $K$-$\Gamma$, and $H$-$A$ paths.

We focus on the crossings indicated by the arrows in Fig.~\ref{fig:band}(a) and scan the BZ to trace the nodal structure. We find that these points are in fact located on three groups of nodal lines, as depicted in Fig.~\ref{fig:nodal}(b). Group $a$ contains two nodal lines pinned along the $K$-$H$ and $K'$-$H'$ paths; Group $b$ contains six nodal lines constrained in the three vertical mirror planes; and Group $c$ has three members lying in the horizontal mirror plane, each forming a ring around the $M$ point. Here, each of the nodal lines carries a topological charge given by the quantized $\pi$ Berry phase $\nu=\oint_C \text{Tr}\bm{\mathcal{A}}\cdot d\bm k\, \mod 2\pi$, where $C$ is a closed path encircling the line, $\bm{\mathcal{A}}$ is the Berry connection for the occupied bands, and the $\pi$ quantization is enforced by the $\mathcal{PT}$ symmetry in the absence of SOC. Perturbations that respect the $\mathcal{PT}$ symmetry may deform the shapes of the nodal lines but cannot destroy them. Including SOC can open a small gap ($\sim$ 20 meV) at these nodal lines.



{\color{blue}\textit{Discussion.}} Our DFT+DMFT study indicates that  KV$_3$Sb$_5$ family materials are good metals with weak electronic correlation effects. The mass enhancement is only around $1.3$ to $1.4$. The obtained Pauli-like paramagnetism and absence of local moment are consistent with recent experimental results. The revealed nodal line structures could be probed by ARPES. Due to the $\pi$ Berry phase of the nodal lines, we expect a suppressed back scattering for the in-plane transport, which is an important factor underlying the materials' good conductivity.

It should be noted that our current study can only deal with on-site correlations. Nonlocal correlations are not included. Our result hence implies that local correlations alone are not sufficient for realizing an unconventional superconductivity. If the superconductivity is indeed unconventional, it would hint at an important role played by nonlocal correlations.


\begin{acknowledgments}
The authors thank Gang Xu, Zhi-Guo Chen, M. R. Kim and D. L. Deng for valuable discussions. This work was supported by the National Natural Science Foundation of China (No.~11604273), and the Singapore Ministry of Education AcRF Tier 2 (MOE2019-T2-1-001).
The computational work was performed on resources of the National Supercomputing Centre, Singapore.
\end{acknowledgments}

\bibliography{refs}

\end{document}